\documentstyle[11pt,newpasp,twoside,epsf]{article}
\begin{document}
\title{The Statistics of Extrasolar Planets: Results from the Keck
  Survey} 
\author{Andrew Cumming\altaffilmark{1}, Geoffrey W. Marcy\altaffilmark{2}, R. Paul Butler\altaffilmark{3}, \& Steven S. Vogt\altaffilmark{1}}\affil{\altaffilmark{1}UCO/Lick Observatory and Department of Astronomy and Astrophysics, University of California, Santa Cruz, CA 95064}
\affil{\altaffilmark{2}Dept. of Astronomy, University of California, Berkeley, CA 94720}
\affil{\altaffilmark{3}Dept. of Terrestrial Magnetism, Carnegie Institution of Washington, 5241 Broad Branch Road
NW, Washington, DC 20015-1305}

\begin{abstract}
We present an analysis of precision radial velocity measurements for
580 stars from the Keck survey. We first discuss the detection
threshold of the survey, and then describe a Bayesian approach to
constrain the distribution of extrasolar planet orbital parameters
using both detections and upper limits.
\end{abstract}

\section{Introduction}
Attempts to characterize the distribution of mass, orbital radius, and
eccentricity of extrasolar planets are hampered by the lack of
knowledge of detection sensitivities. Following our previous study of
the Lick survey (Cumming, Marcy, \& Butler 1999, hereafter CMB), the
aim of this work is to carefully assess the detection threshold and
selection biases for a sample of stars from the Keck survey, and use
this information to constrain the distribution of orbital parameters.

\section{Search for Companions}

We analyse radial velocity measurements for 580 F,G,K,M stars from the
Keck survey. Typically, each star has 3 or 4 measurements per year
over 2 to 5 years with measurement errors between 2 and 5 m/s.

We adopt a least-squares approach. First, we check whether the
velocity scatter is consistent with measurement errors and intrinsic
stellar ``jitter''. The jitter, which arises from a combination of
convective motions, magnetic activity, and rotation, is estimated from
stellar parameters such as rotation period and mass. Assuming Gaussian
statistics, we calculate the probability of obtaining $\chi^2$ as
large or larger than the observed value from measurement errors and
jitter alone, adopting a threshold probability of $10^{-3}$. The solid
line in Figure 1 shows the signal to noise needed for a signal to be
detected by this method 50\% of the time as a function of the number
of data points $n$. We define the signal to noise as
$K/\sqrt{2}\sigma$, where $K$ is the velocity amplitude, and $\sigma$
the expected noise amplitude. Even with only a few measurements,
signals $>2\sigma$ are detected.

Next, we look for periodicities using the Lomb-Scargle periodogram
(Scargle 1982; Walker et al.~1995; CMB), which is similar to the
F-statistic, measuring the improvement in $\chi^2$ between a model
which includes a sinusoid and one that does not. We look for the
maximum periodogram power between 2 days and 10 years, and use Monte
Carlo methods to assess its significance. The dashed lines (analytic
\begin{figure}
\begin{center}
\epsfxsize=3.75in
\epsfbox{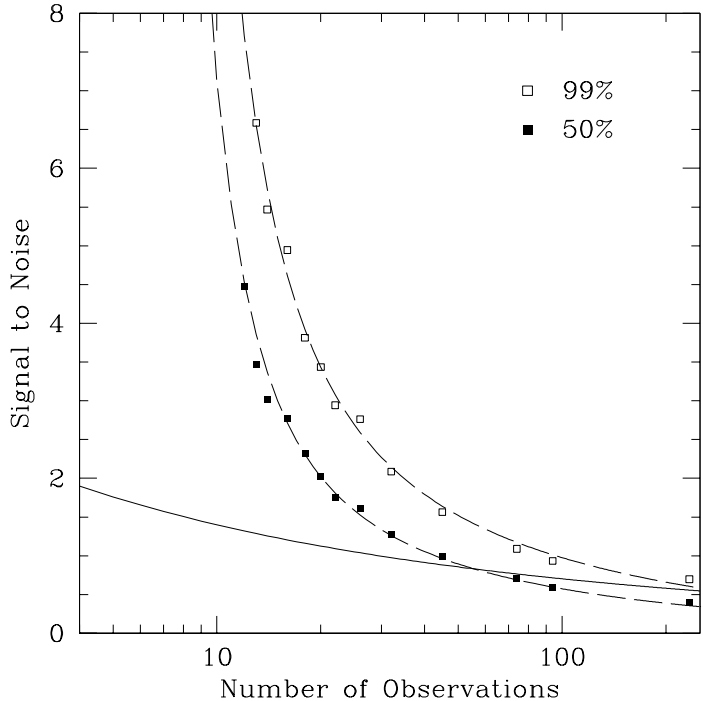}
\end{center}
\vspace{-0.5cm}
\caption{Signal to noise ratio $K/\sqrt{2}\sigma$ needed for detection
by $\chi^2$-test (solid line) or periodogram (dashed lines). For small
$n$, the dashed lines increase much more rapidly than $1/\sqrt{n}$.}
\end{figure}
result) and squares (numerical result) in Figure 1 show the signal to
noise ratio needed to identify a periodic signal 50\% (filled squares)
and 90\% (open squares) of the time (significance level $10^{-3}$) as
a function of number of data points. To obtain analytic estimates, we
follow the approach of Horne \& Baliunas (1986), but using the correct
statistical distribution for periodogram powers in the presence of
noise (CMB Appendix B). We find the signal to noise ratio needed for
detection 50\% of the time is
$K/\sqrt{2}\sigma=[(M/F)^{2/(n-3)}-1]^{1/2}$, where $M$ the number of
independent frequencies searched, and $F$ the false alarm probability
needed for detection. For large $n$ the 50\% threshold is
$K/\sqrt{2}\sigma\sim[2\ln(M/F)/n]^{1/2}$, or $\approx 5/\sqrt{n}$ for
typical values $F=10^{-3}$ and $M=1000$.

Figure 1 is striking because when looking for periodic signals, we are
used to being able to ``dig into the noise''. However, here we are
limited by the small number of observations. Points to note are: (i)
even with small $n$, we detect excess variability if $K> 1$--$2\
\sigma$, (ii) it is hard to characterize an orbit when the number of
observations is $\leq 10$, (iii) we detect periodic signals with
$K\approx 2$--$4\ \sigma$ when $n=10$--$20$, and (iv) detecting
amplitudes $<1\ \sigma$ requires $n>50$.

\begin{figure}
\begin{center}
\epsfxsize=4in
\epsfbox{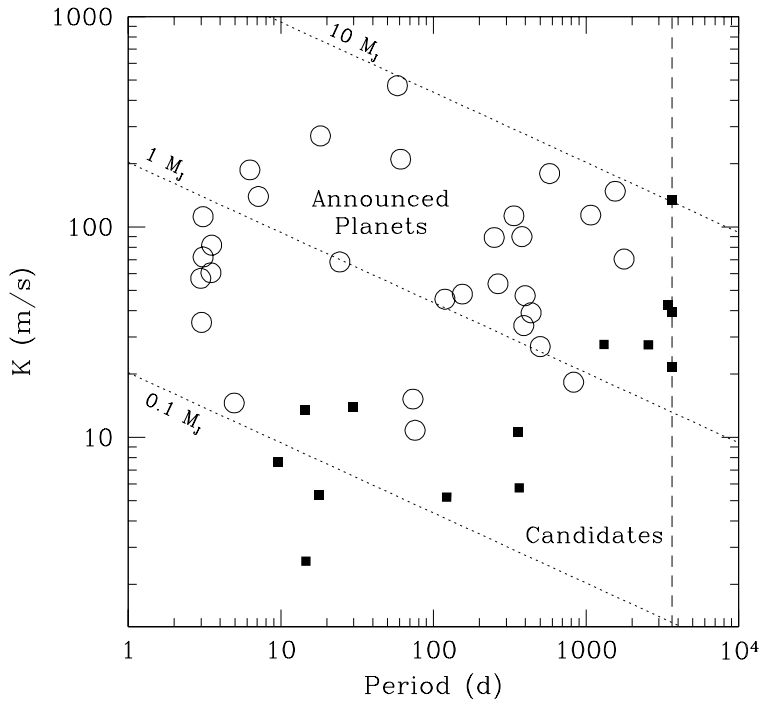}
\end{center}
\vspace{-0.5cm}
\caption{Announced planets (open circles) and candidates (filled
squares) (periodogram false alarm $<10^{-3}$). Sloping dashed lines
show different values of $M\sin i$ for a $1\ M_\odot$ star. Vertical
dashed line shows maximum period searched. We show only fits with
$M\sin i<10 M_J$.}
\end{figure}

We are currently extending this approach to include eccentric
orbits. However for now, we use the best-fit circular orbit amplitude
and period from the periodogram as initial conditions for a full
Keplerian fit.  The resulting Keplerian amplitudes and periods for the
significant detections are shown in Figure 2. Most, but not all, of
the candidates from the periodogram analysis are announced planets
(open circles). Those candidates that have not been announced (filled
squares) have either low amplitude or long orbital period.  There
is an ``effective detection threshold'' beyond that set by the
statistics of the periodogram. This comes from (i) we must be sure the
observed variations are not due to periodic stellar jitter, and (ii)
inability to characterize orbits longer than the duration of the
observations. The announced planets have velocity amplitude
$>10$--$15\ {\rm m/s}$, periods less than the duration of the
observations ($<3$--$5\ {\rm years}$), and number of data points
$n>10$. Better characterization of jitter would aid detection of low
amplitude planets.

Finally, for most stars with no detection, we can exclude companions
with velocity amplitudes $>10$--$20\ {\rm m/s}$, or $M\sin
i>(0.35$--$0.7) M_J\ (a/AU)^{1/2}$, for orbital periods less than the
duration of the observations (2 to 5 years).

\section{Constraining the Distribution of Orbital Parameters}

We adopt a Bayesian approach to constrain the distribution of masses,
orbital periods, and eccentricities {\it using both detections and
upper limits}. These calculations are currently in progress, here we
discuss the methodology. Consider a model of the distribution of
masses and orbital periods $n(M,P)$ (normalized so that $\int n(M,P)
dM dP=1$) and fraction of stars with planets $f$. We write a
likelihood function $L=\prod_j L_j$ where $j$ denotes each star, and
\begin{equation}
L_j=(1-f)q_j+f\int\ dM\ dP\ n(M,P)\ p_j(M,P).
\end{equation}
This expression includes the probability of the data {\it given a
planet} of mass $M$, period $P$,
\begin{equation} 
p_j(M,P)=\int {d(\cos i)}\ \int {d\phi\over 2\pi}\ \prod_{k=1}^{N}\
{1\over \sqrt{2\pi}\sigma_k}\exp\left(-{\left[v_k-f_k(M\sin
i,P,\phi)\right]^2\over 2\sigma_k^2}\right),
\end{equation}
and the probability of the data {\it given no planet} $q_j$. No
decision must be made for each star about whether a planet has been
detected or not: both possibilities are included, weighted by their
relative probability. This approach can be readily generalized to
include, for example, eccentric orbits.

As a check, imagine we can definitely say whether a star has a planet
or not. The likelihood is then $L=f^{N_p}(1-f)^{N_\star-N_p}$ where
$N_p$ planets are detected out of $N_\star$ stars. Maximizing $L$ with
respect to $f$ gives the expected result $f=N_p/N_\star$, the best
guess at the fraction of stars with planets is the observed fraction.

Two striking features already apparent are (i) the lack of massive
planets at short periods (easiest to detect, but not seen), and (ii)
the ``pile-up'' of planets at orbital periods $\approx 3\ {\rm days}$
(no planets were found between $2$ and $3$ days in our search,
suggesting that this is not a selection effect). We are currently
investigating these issues and others more carefully, including (i)
the fraction of stars with planets, (ii) whether there is a lack of
Saturn-mass companions at short orbital periods, and (iii) whether
planet properties depend on the spectral type of the host star.

\acknowledgements We thank G. Ushomirsky, D. Chernoff, I. Wasserman,
R. Rutledge and D. Reichart for useful discussions. AC thanks Caltech
Astronomy Department for hospitality during a recent visit. This work
was supported by NASA Hubble Fellowship grant HF-01138 awarded by the
Space Telescope Science Institute, which is operated for NASA by the
Association of Universities for Research in Astronomy, Inc., under
contract NAS 5-26555.


\begin{references}

\noindent
Cumming, A., Marcy, G. M., \& Butler, R. P. 1999, ApJ, 526, 890 (CMB)

\noindent
Horne, J. H., \& Baliunas, S. L. 1986, ApJ, 302, 757

\noindent
Scargle, J. D. 1982, ApJ, 263, 835

\noindent
Walker, G. A. H., Walker, A. R., Irwin, A. W., Larson, A. M., \& Yang,
S. L. S. 1995, Icarus, 116, 359

\end{references}
\end{document}